\documentclass[review]{elsarticle}

\usepackage{lineno,hyperref}
\usepackage{color}
\usepackage{amsmath}
\usepackage{amsfonts}
\usepackage{amssymb}
\usepackage{amsbsy}
\usepackage{amsthm}
\usepackage{amssymb}
\usepackage[acronym]{glossaries}
\usepackage{algorithm}
\usepackage[noend]{algpseudocode}
\usepackage{indentfirst} 

\theoremstyle{definition}

\numberwithin{equation}{section}

\def\*#1{\mathbf{#1}}
\def\~#1{\boldsymbol{#1}}

\modulolinenumbers[5]

\biboptions{authoryear}

\begin{document}


\title{Does the leverage effect affect the return distribution?}

\author{Dangxing Chen}
\date{}
\address{Consortium for Data Analytics in Risk, Department of Economics, 530 Evans Hall, University of California, Berkeley, CA, 94720-3880, USA}
\ead{dangxing@berkeley.edu}
 \fntext[myfootnote]{This work was supported by Southwest University of Finance and Economics through the Consortium for Data Analytics in Risk.}



\begin{abstract}
\noindent The leverage effect refers to the generally negative correlation between the return of an asset and the changes in its volatility. There is broad agreement in the literature that the effect should be present for theoretical reasons, and it has been consistently found in empirical work. However, a few papers have pointed out a puzzle: the return distributions of many assets do not appear to be affected by the leverage effect.  We analyze the determinants of the return distribution and find that the impact of the leverage effect comes primarily from an interaction between the leverage effect and the mean-reversion effect. When the leverage effect is large and the mean-reversion effect is small, then the interaction exerts a strong effect on the return distribution.  However, if the mean-reversion effect is large, even a large leverage effect has little effect on the return distribution.  To better understand the impact of the interaction effect, we propose an indirect method to measure it. We apply our methodology to empirical data and find that the S$\&$P 500 data exhibits a weak interaction effect, and consequently its returns distribution is little impacted by the leverage effect. Furthermore, the interaction effect is closely related to the size factor: small firms tend to have a strong interaction effect and large firms tend to have a weak interaction effect. 
\end{abstract}

\maketitle




\section{Introduction}

The leverage effect refers to the observed tendency of changes in an asset's volatility to be negatively correlated with the asset's returns. The original interpretation goes back to \cite{black1976studies}. The decline of the stock increases the financial leverage and thereby increases the volatility.  Conversely, increases in volatility must be compensated by increases in expected future returns, which can only be achieved by lowering the current stock price. There are many subsequent papers examining the cause of the leverage effect (see, e.g.,  \citealp{christie1982stochastic,figlewski2000leverage,french1987expected,campbell1992no}).

Whatever the cause(s) of the leverage effect, there is broad agreement in the literature that the effect is present.  Many papers have attempted to accurately estimate the leverage effect (see, e.g.,   \citealp{wang2014estimation,bandi2012time,yu2005leverage}). For example, \cite{ait2013leverage} calculated that the correlation parameter at the high-frequency limit is $\rho=-0,77$ by high-frequency data of S$\&$P 500 data with a robust estimation, indicating that there is a strong leverage effect. However, when stochastic volatility models (SVMs) are fitted to the S$\&$P 500, a puzzle arises: the fitting is insensitive to the correlation parameter (see, e.g., \citealp{chorro2018testing,dragulescu2002probability,sepp2008pricing}).  How can it be that the return distribution is insensitive to the correlation parameter?  This puzzle is the focus of this paper. 

Our studies rely on the framework of the continuous-time (CT) SVM. The CT-SVM has been widely successful  under both the risk-neutral measure (see, e.g., \citealp{sepp2008pricing,heston1993closed,forde2009small,ahn1999parametric,ait2002maximum}) and the physical measure (see, e.g., \citealp{dragulescu2002probability,silva2003comparison,bakshi2006estimation}). The main advantages are that the CT-SVM is mathematically well-defined and fits the empirical data very well. There are also many useful discrete-time (DT) SVM (see, e.g., \citealp{bauwens2006multivariate,duan1995garch,lamoureux1990heteroskedasticity}). 
In particular, the DT-SVM EGARCH model derived by \cite{nelson1991conditional} is widely used in practice. However, the CT-SVM usually has a more flexible form than the DT-SVM, and the discretization from continuous time to discrete time doesn't seem always appropriate. For instance, in empirical data, the annualized marginal variance may grow over time; this is not allowed in the EGARCH model. Therefore, in this paper, we would only focus on the CT-SVM.

Under our framework, using the stochastic volatility model of \cite{heston1993closed} as an example, we see that the leverage effect is important, but it need not  have a strong impact on the return distribution. In particular, the impact of a strong leverage effect on the return distribution can be negated by a strong mean-reversion effect. Hence, when studying the distribution properties, one must consider the interaction effect between the leverage effect and the mean-reversion effect. The interaction effect is not directly measurable, since volatility is latent and the Brownian motion is not observable. We propose an indirect method, relying on calculating the dynamics of the marginal variance. A direct impact of this measurement is that the annualized marginal variance will grow over time for a strong interaction effect, but barely change for a weak interaction effect. This phenomenon also provides a better understanding of the performance of the square-root-of-time rule (SRTR) (see, e.g., \citealp{danielsson2006time,wang2011accurate,chen2018predicting}). It is well known that, in the presence of a strong mean-reversion effect, SRTR underpredicts annual volatility when current volatility is low, and overpredicts annual volatility when current volatility is high.  We show that, even with a weak mean-reversion effect, if the leverage effect is strong, then the interaction effect has a strong impact on the return distribution. As a result, the SRTR tends to underpredict annual volatility on average. 

We apply our methodology to study the relationship between the interaction effect with firm size. The empirical evidence indicates that the interaction effect is strong for small firms but weak for large firms. As a consequence, the annualized marginal variance grows over time for small firms but barely change for large firms. 

Finally, we aim to explain the leverage effect puzzle in S$\&$P 500. By using the generalized hyperbolic distribution (see \citealp{eberlein1995hyperbolic}), a distribution that arises from the CT-SVM when $\rho=0$, fits the S$\&$P 500 return distribution very well, despite the fact that the S$\&$P 500 exhibits a strong leverage effect.  By our method, we show that the interaction effect of the S$\&$P 500 is weak, answering the puzzle.

The paper is organized as follows. In Section 2, we introduce our basic framework of the stochastic volatility model, followed with a detailed example, the Heston model. Section 3 documents the presence of the leverage effect puzzle, and provides the explanation and the solution.  The relationship of the interaction effect with firm size is explored in Section 4. Section 5 studies the leverage effect puzzle in S$\&$P 500. Section 6 concludes.

\section{Stochastic volatility model}
We assume the price of a security follows the stochastic differential equation (SDE)
\begin{align}
dS_t &= r S_t \ dt + \sqrt{V_t} S_t \ d \widetilde{B}_t, \label{eq:price_dynamic} \\
dV_t &= \mu(V_t) \ dt + \sigma(V_t) \ dW_t. \label{eq:var_dynamic}
\end{align}
where $\widetilde{B}_t$ and $W_t$ are two standard Brownian motions with $\mathbb{E}[d \widetilde{B}_t dW_t] = \rho \ dt$, and $r$ is the rate of return.\footnote{Estimates of the mean rate of return are notoriously noisy even over periods of years or decades. Since we cannot make meaningful estimates of $r_t$, we might as well assume it as constant.}  Note that 
\begin{align}
\rho = \lim_{s \rightarrow 0} \ \text{Corr}(V_{t+s}-V_t, X_{t+s}-X_t)
\end{align}
so that the leverage effect is summarized by the correlation parameter $\rho$ under the model \eqref{eq:price_dynamic} and \eqref{eq:var_dynamic}.  It is convenient to apply the  Gram-Schmidt process to rewrite the price dynamics in terms of two independent Brownian motions
\begin{align}
dS_t = rS_t \ dt + \rho \sqrt{V_t} S_t \ dW_t + \sqrt{1-\rho^2} \sqrt{V_t} S_t \ dB_t.
\end{align} 
The log price dynamics $X_t = \ln(S_t)$ can be derived by It\^{o}'s Lemma,
\begin{align} \label{eq:return_dynamic}
dX_t = \left( r - \frac{1}{2} V_t \right) \ dt + \rho \sqrt{V_t} \ dW_t + \sqrt{1-\rho^2} \sqrt{V_t} \ dB_t.
\end{align}
Notice that, for greater generality, we don't specify the detailed form of the variance process. The choice of $\mu(V_t)$ and $\sigma(V_t)$ can be quite flexible. We assume that the initial variance $V_0>0$ is a realization from the stationary (invariant) distribution of \eqref{eq:var_dynamic}, so that $V_t$ is a stationary process. 

We will focus on the marginal return distribution, since it can be directly estimated from the empirical data. We derive some asymptotic properties of the marginal return distribution. We consider the centralized and scaled return distribution
\begin{align}
\widetilde{X}_t = \frac{X_t - \mathbb{E}[X_t]}{\sqrt{t}}.
\end{align}
At a short horizon $t \rightarrow 0$, we have
\begin{align}
\widetilde{X}_t | V_0 \rightarrow \mathcal{N} \left( 0, V_0 \right).
\end{align}
With this expression, the related moments can be calculated 
\begin{align}
\text{Var}[\widetilde{X}_t] &= \mathbb{E}[V_0], \\
\text{Skewness}[\widetilde{X}_t] &= 0, \\
\text{Kurtosis}[\widetilde{X}_t] &= 3 + 3 \frac{\text{Var}[V_0]}{\mathbb{E}^2[V_0]}. \label{eq:return_asymp_kur}
\end{align}
Empirically, as one increases the time scale over which returns are calculated, their distributions looks more and more like a Gaussian distribution, as discussed by \cite{cont2001empirical} .  At a long horizon $t \rightarrow \infty$, under some mild conditions (see \citealp{peligrad1986recent}) , which we believe is satisfied empirically, we have
\begin{align} \label{eq:return_long}
\widetilde{X}_t \rightarrow \mathcal{N} \left( 0, \frac{\text{Var}[X_t]}{t} \right).
\end{align}
The central moments of the Gaussian distribution are well known:\footnote{$(p-1)!!$ denotes the double factorial, i.e. the product of all odd numbers $1, \ldots, (p-1)$.}
\begin{align*}
\mathbb{E}[\widetilde{X}_t^p] = 
\begin{cases}
& 0, \ \ \ \text{if $p$ is odd}, \\
& \left( \frac{\text{Var}[X_t]}{t} \right)^{p/2}(p-1)!!,  \ \ \ \text{if $p$ is even}.
\end{cases}
\end{align*}

\subsection{Heston model}
Throughout the paper, we will frequently use the stochastic volatility model of \cite{heston1993closed}, which has the advantage of providing explicit expressions for many distributional properties. The equation is written as
\begin{align}
dX_t &= \left( r - \frac{1}{2}V_t \right) \ dt + \rho \sqrt{V_t} \ dW_t + \sqrt{1-\rho^2} \sqrt{V_t} \ dB_t, \\
dV_t &= \kappa(\theta-V_t) \ dt + \sigma \sqrt{V_t} \ dW_t.
\end{align}
The dynamic of the variance process in the Heston model is also known as the Cox-Ingersoll-Ross (CIR) process (see \citealp{cox2005theory}). We assume in what follows that the Feller condition $2\kappa \theta > \sigma^2$ holds, which guarantees that the variance process is always strictly positive.

The marginal density function can be written in terms of the Fourier integral (see \citealp{dragulescu2002probability})
\begin{align} \label{eq:Heston_density}
P_t(x) = \frac{1}{2\pi} \int_{-\infty}^{+\infty} e^{ip_x x + F_t(p_x)} \ dp_x,
\end{align}
with
\begin{align}
F_t(p_x) &= \frac{\kappa \theta}{\sigma^2} \Gamma t - \frac{2\kappa \theta}{\sigma^2} \ln \left[ \text{cosh} \left( \frac{\Omega t}{2} \right) + \frac{\Omega^2-\Gamma^2+2\kappa \Gamma}{2\kappa \Gamma} \text{sinh} \left( \frac{\Omega t}{2} \right) \right], \\
\Gamma &= \kappa + i \rho \sigma p_x, \\
\Omega &= \sqrt{\Gamma^2 + \sigma^2(p_x^2-ip_x)}.
\end{align}
With this, the marginal density function of return can be recovered via the fast Fourier transform (see, e.g., \citealp{valsa1998approximate,abate1995numerical})

Many formulas regarding the marginal moments can then be derived from Formula \eqref{eq:Heston_density}. For an intuitive understanding, we consider the first four moments. The marginal expectation is given by
\begin{align}
\mathbb{E}[X_t] = rt - \frac{\theta t}{2}.
\end{align}
The marginal variance is given by
\begin{align}
\text{Var}[X_t] = \mathbb{E} \left[ \int_0^t V_s \ ds \right] + \frac{1}{4} \text{Var} \left[ \int_0^t V_s \ ds \right] - \rho \mathbb{E} \left[ \left( \int_0^t V_s \ ds \right) \left( \int_0^t \sqrt{V_s} \ dW_s \right) \right].
\end{align}
and the expression for these terms are
\begin{align} 
\mathbb{E} \left[ \int_0^t V_s \ ds \right] &= \theta t, \label{eq:Heston_EIV} \\
\mathbb{E} \left[ \left( \int_0^t V_s \ ds \right) \left( \int_0^t \sqrt{V_s} \ dW_s \right) \right] &= \theta \frac{\sigma}{\kappa} \left[t+ \frac{e^{-\kappa t}-1 }{\kappa} \right], \label{eq:Heston_EMIV} \\
 \text{Var} \left[ \int_0^t V_s \ ds \right] &= \theta \frac{\sigma^2}{\kappa^2} \left[ t + \frac{e^{-\kappa t}-1}{\kappa} \right]. \label{eq:Heston_VIV}
\end{align}
The marginal skewness is 
\begin{align}
\text{Skewness}[X_t] = \frac{\mathbb{E}[(X_t-\mathbb{E}[X_t])^3]}{\text{Var}[X_t]^{3/2}},
\end{align}
where
\begin{align}
\mathbb{E}[(X_t-\mathbb{E}[X_t])^3] &= \frac{3}{8\kappa^5} e^{-\kappa t} \theta \sigma (-2\kappa \rho + \sigma) ( -4\kappa^2+8 \kappa \rho \sigma + 4t \kappa^2 \rho \sigma - 2\sigma^2 - t \kappa \sigma^2 \\
&+ e^{\kappa t}(-4 \kappa^2(-1+t\kappa) + 4\kappa (-2+t\kappa) \rho \sigma + (2-t\kappa)\sigma^2) ).
\end{align}
The marginal kurtosis is 
\begin{align}
\text{Kurtosis}[X_t] = \frac{\mathbb{E}[(X_t-\mathbb{E}[X_t])^4]}{\text{Var}[X_t]^2},
\end{align}
where
\begin{align*}
\mathbb{E}[(X_t-\mathbb{E}[X_t])^4] &= \frac{3}{32\kappa^7} e^{-2\kappa t} \theta (\sigma^2(-4\kappa \rho + \sigma)^2(2\theta \kappa+\sigma^2) \\
&+ 4e^{\kappa t} \sigma (-16t \theta \kappa^5 \rho + 4 \kappa^3((2+t\theta)\kappa + 4(\theta(-1+t\kappa)+2\kappa(2+t\kappa))\rho^2)\sigma \\
&-8\kappa^2 \rho (\theta(-1+t\kappa)+6\kappa(2+t\kappa)+2\kappa(6+t\kappa(4+t\kappa))\rho^2)\sigma^2 \\
&+\kappa(\theta(-1+t\kappa)+4\kappa(6+3t\kappa+(34+t\kappa(24+5t\kappa))\rho^2))\sigma^3 \\
&-8\kappa(7+t\kappa(5+t\kappa))\rho\sigma^4+(7+t\kappa(5+t\kappa))\sigma^5) \\
&+ e^{2\kappa t}(2t^2\theta \kappa^3 (4\kappa^2-4\kappa \rho \sigma + \sigma^2)^2 \\
&+ 2t \kappa \sigma (4\kappa^2-4\kappa \rho \sigma+\sigma^2)(8\theta \kappa^2 \rho + 2\kappa(-\theta+2\kappa+8\kappa \rho^2)\sigma - 20\kappa \rho \sigma^2 + 5\sigma^3) \\
&+ \sigma^2(-32\kappa^4(1+8\rho^2)-29\sigma^4 + 2\kappa \sigma^2(\theta+116 \rho \sigma) + 32 \kappa^3 \rho (\theta \rho + 12(1+\rho^2)\sigma) \\
&- 16 \kappa^2 \sigma (6\sigma + \rho (\theta + 35 \rho \sigma))))). 
\end{align*}

\section{The interaction effect}

\subsection{The leverage effect puzzle}
\begin{figure}[H]
\centering
\includegraphics[scale=0.6]{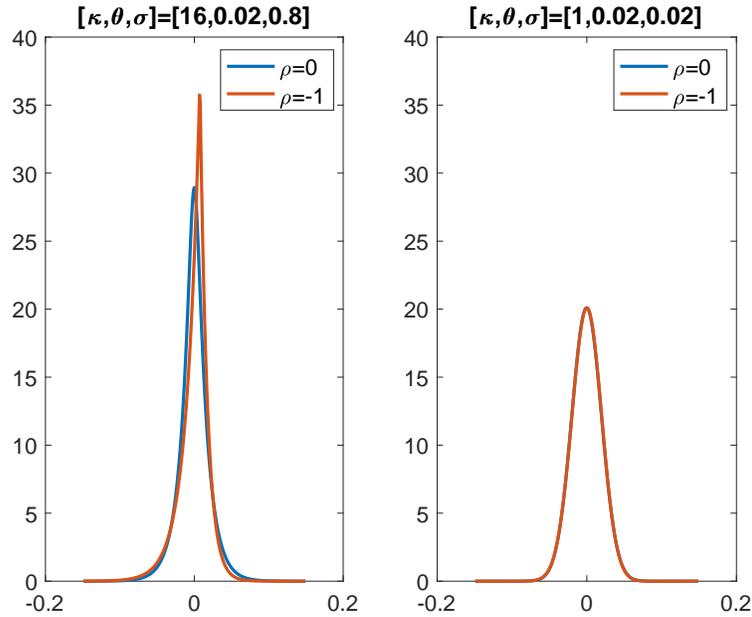}
\caption{The marginal density of the Heston model at 5-day horizon with and without the leverage effect}
\label{fig:Heston_distn}
\end{figure}

To motivate the analysis that follows, we start with a straightforward artificial example to illustrate the leverage effect puzzle. Figure~\ref{fig:Heston_distn} compares the marginal densities of the return in the Heston model \eqref{eq:Heston_density} with and without the leverage effect for two sets of Heston parameters $\kappa, \sigma$.  In the first example with $\kappa=16$ and $\sigma=0.8$, we see a strong impact of the leverage effect. In fact, with $\rho=-1$, the marginal density is quite negatively skewed.  In the second example, with $\kappa =1$ and $\sigma=0.02$, we see that the two marginal densities with $\rho=0$ and $\rho=-1$ are almost identical. So why does the impact of the leverage effect almost disappear in some cases? That is the ``leverage effect puzzle" that we seek to understand. The goal of this paper is to understand the sources of the puzzle and propose a solution. 

\subsection{Interaction effect}

When the leverage effect is not present, i.e., $\rho=0$, the solution to the SDE \eqref{eq:return_dynamic} can be written as
\begin{align} \label{eq:return_cond_indep}
X_t \bigg| \int_0^t V_s \ ds = \mathcal{N} \left( r - \frac{1}{2} \int_0^t V_s \ ds, \int_0^t V_s \ ds \right).
\end{align}
Now consider another case when there is a strong mean-reversion effect such that $V_t$ is independent of $W_t$, then the equation \eqref{eq:return_cond_indep} also holds and the impact of the leverage effect is not observed. This situation could  occurs, for example, when the effect from the diffusion term is relatively weak. Returning to the example of the Heston model in Figure~\ref{fig:Heston_distn}, let's see the impact of the leverage effect on the marginal moments of return. At a short horizon $t \sim 0$, the following simple approximations follow from equations \eqref{eq:Heston_EIV}, \eqref{eq:Heston_EMIV}, and \eqref{eq:Heston_VIV} 
\begin{align*}
\mathbb{E}[X_t] &\sim rt - \frac{\theta t}{2}, \\
\text{Var}[X_t] &\sim \theta t, \\
\text{Skewness}[X_t] &\sim \sigma \left( \frac{3}{2} \rho - \frac{3}{4} \frac{\sigma}{\kappa} \right) \sqrt{\frac{t}{\theta}}, \\
\text{Kurtosis}[X_t] &\sim 3 + \frac{3\sigma^2}{2\kappa \theta}.
\end{align*}
Note the expectation, variance, and kurtosis are not affected by the leverage effect, but the skewness is. Even with a strong leverage effect, if $\sigma$ is small (i.e. the variance process $V_t$ is not very volatile), the skewness will be close to 0 and the impact of the leverage effect will be hard to observe.

Hence, a strong leverage effect can be negated by a strong mean-reversion effect.  When studying return distributions, it is essential to incorporate the interaction effect, namely the interaction between the mean-reversion and leverage effects. Since the variance is latent and the Brownian motion is not observable, we propose an indirect method to measure the interaction effect.

\subsection{Measurement of the interaction effect}

We need an accurate, robust, and interpretable measure of the interaction effect. Commonly used measurements in summary statistics include the location (e.g., mean, median),  dispersion (e.g., standard deviation), and shape (skewness and kurtosis). For these measurements, calculations typically rely on moments or quantiles.  It would be natural to focus on the moment-based calculations, since these can be readily interpreted in terms of the SDE.  Unfortunately, the estimation of moments in financial data may be inaccurate.  For example, consider the expectation of return $\mathbb{E}[X_t] = rt - \frac{\theta t}{2}$. At a short horizon, the standard deviation of the return $\sqrt{\text{Var}[X_t]} \sim \sqrt{\theta t}$ has a much larger magnitude than its expectation. As a consequence, we cannot accurately estimate the mean return even with about 100 years of daily return, due to the slow convergence under the central limit theorem.  The lack of robustness in the calculation of skewness and kurtosis in financial data using moments is documented in \cite{kim2004more} and \cite{bonato2011robust}. 

Therefore, among these common measurements, we focus on the calculation of the marginal variance since it can be estimated accurately and is easily interpretable.  Recall that by algebra, the marginal variance can be decomposed into three terms: 
\begin{align}
\text{Var}[X_t] = \mathbb{E} \left[ \int_0^t V_s \ ds \right] + \frac{1}{4} \text{Var} \left[ \int_0^t V_s \ ds \right] - \rho \mathbb{E} \left[ \left( \int_0^t V_s \ ds \right) \left( \int_0^t \sqrt{V_s} \ dW_s \right) \right].
\end{align}
For simplicity, we use the following notation for these terms:
\begin{align}
\text{EIV}_t &= \mathbb{E} \left[ \int_0^t V_s \ ds \right], \\
\text{VIV}_t &= \frac{1}{4} \text{Var} \left[ \int_0^t V_s \ ds \right], \\
\text{EMIV}_t &= - \rho \mathbb{E} \left[ \left( \int_0^t V_s \ ds \right) \left( \int_0^t \sqrt{V_s} \ dW_s \right) \right].
\end{align}
Note that the $\text{EIV}_t$ is scalable with respect to time, i.e., $\text{EIV}_t = \frac{t}{s} \text{EIV}_s$. Due to the mean-reversion effect, empirically, we believe the $\text{VIV}_t$ is negligible comparing to the $\text{EIV}_t$, i.e., $\text{VIV}_t \ll \text{EIV}_t$. Then the $\text{EMIV}_t$ comes from the contribution from the interaction effect and is the term we want to measure. The $\text{EMIV}_t$ should be close to 0 for a weak interaction effect, far away from 0 for a strong interaction effect.

For a more intuitive understanding of these formulas, we use the Heston model as an example, taking formulas from Equations \eqref{eq:Heston_EIV}, \eqref{eq:Heston_EMIV}, and \eqref{eq:Heston_VIV}. At a short horizon $t\sim 0$ , we have
\begin{align}
\mathbb{E} \left[ \int_0^t V_s \ ds \right] &\sim \theta t, \\
 \mathbb{E} \left[ \left( \int_0^t V_s \ ds \right) \left( \int_0^t \sqrt{V_s} \ dW_s \right) \right] &\sim  \theta \frac{\sigma}{2} t^2, \\
\text{Var} \left[ \int_0^t V_s \ ds \right] &\sim \theta \frac{\sigma^2}{2\kappa} t^2.
\end{align}
Thus, at a short horizon $t\sim 0$, we have $\text{Var}[X_t] \sim \text{EIV}_t$. At a long horizon $t \sim \infty$, we have
\begin{align}
\mathbb{E} \left[ \int_0^t V_s \ ds \right] &\sim \theta t, \\
\mathbb{E} \left[ \left( \int_0^t V_s \ ds \right) \left( \int_0^t \sqrt{V_s} \ dW_s \right) \right] &\sim \theta t \frac{\sigma}{\kappa}, \\
\text{Var} \left[ \int_0^t V_s \ ds \right] &\sim \theta t \frac{\sigma^2}{\kappa^2}.
\end{align}
Thus, the contribution from the $\text{EMIV}_t$ can be observed. In practice, the calculation is done by $\text{EMIV}_t \approx \text{Var}[X_t] - \frac{t}{s} \text{Var}[X_s]$ for $t>s$.
Note that each term is scaled by a factor $\frac{\sigma}{\kappa}$. This term serves as a mean-reversion factor in the Heston model.

From the calculation of the interaction effect by $\text{EMIV}_t$, there is a direct impact: the annualized marginal variance will grow over time until it converges when the interaction effect is strong, but it will barely change when the interaction effect is weak. This phenomenon suggests that in practice, if one observes two securities with the same average
variance at a short horizon, it doesn't imply that these two securities have the same volatility. If the interaction effect is strong for one and weak for another one, the security with the strong interaction effect will become more volatile over time.

Another advantage of measuring the marginal variance is we know the centralized and scaled return $\widetilde{X}_t$ would eventually converge to Gaussian \eqref{eq:return_long}. Suppose the information of marginal variance is known, we then know the asymptotic distribution of the return. In practice, the asymptotic distribution of the return can be easily obtained for a security with a weak interaction effect.

\subsection{Impact on the square-root of time rule}

\begin{figure}[H]
\centering
\includegraphics[scale=0.25]{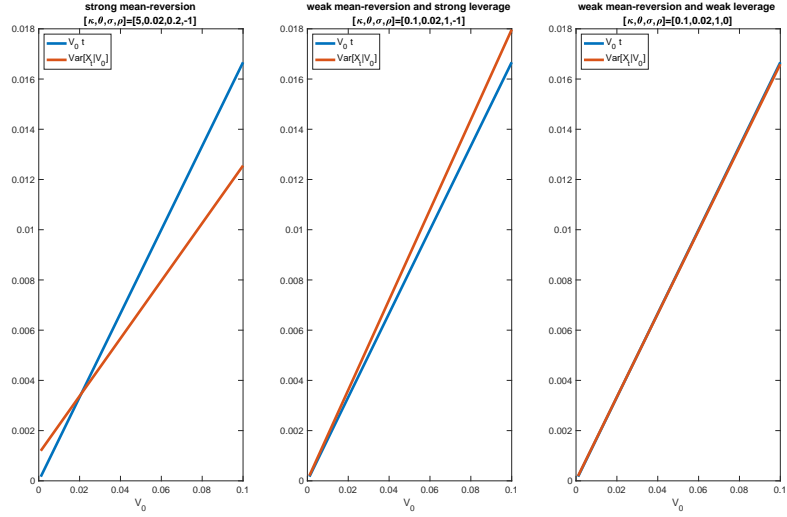}
\caption{Performance of the square-root-of-time rule by Heston model for different cases at 2-month horizon}
\label{fig:Heston_sqrt}
\end{figure}

It is common to use the SRTR (see \citealp{danielsson2006time}) to extrapolate the conditional variance of the log return $\text{Var}[X_t|V_0]$. Under the SVM, the SRTR serves as the constant approximation to the conditional variance
\begin{align*}
\text{Var}[X_t|V_0] \approx V_0 t.
\end{align*}
We discuss the impact of the mean-reversion effect and the leverage effect on the performance of the SRTR, along with the Heston model as an example in Figure~\ref{fig:Heston_sqrt}. It is well known that, because it ignores the mean-reversion effect,  the SRTR tends to under-predict for small initial variance but over-predict for large initial variance.  As shown in the first example in Figure~\ref{fig:Heston_sqrt}, due to the mean-reversion effect, the conditional variance is flatter than the SRTR. But this is not the whole story; one must also take the interaction effect into account.  If the mean-reversion effect is weak and the leverage effect is strong, the resulting strong interaction effect leads SRTR tends to under-predict on average, as shown in the second example of Figure~\ref{fig:Heston_sqrt}. This also explains the downward-biased prediction using the SRTR observed by \cite{wang2011accurate}. In the case of a weak mean-reversion effect and a weak leverage effect, the SRTR serves as a good approximation, as plotted in the third example in Figure~\ref{fig:Heston_sqrt}.

\section{Relationship of the interaction effect with firm size}

Here, we use five value-weighted size portfolios constructed by Fama and French\footnote{The data is downloaded from the website http://mba.tuck.dartmouth.edu/pages/faculty/ken.french/data$\_$library.html}, sorted by market capitalization, price times shares outstanding. All NYSE, AMEX, and NASDAQ stocks are included. We use daily data from July 1926 to January 2019.

\begin{figure}[H]
\centering
\includegraphics[scale=0.4]{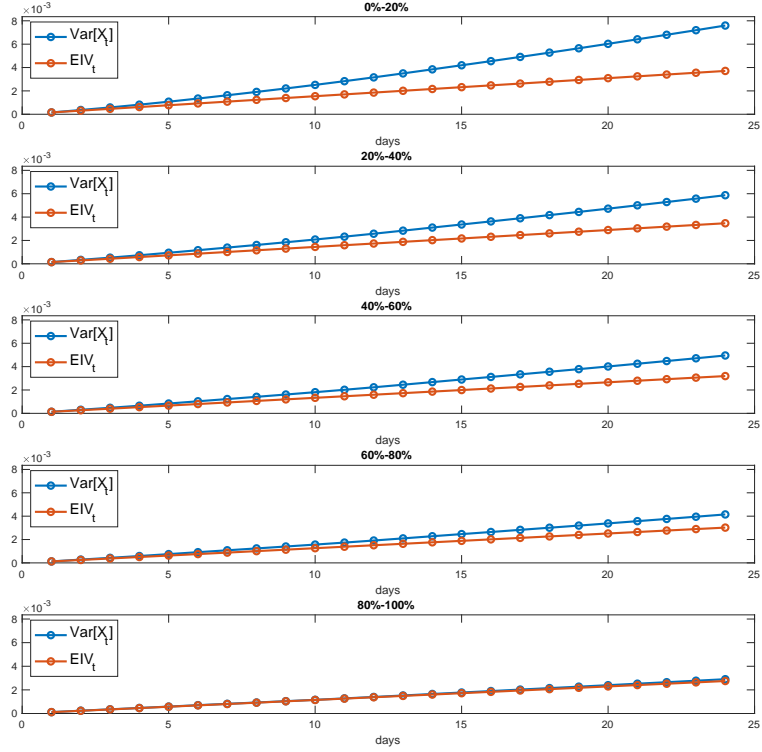}
\caption{The interaction effect for portfolios constructed by size}
\label{fig:size_leverage}
\end{figure}

To study the interaction effect, we compare the dynamics of the marginal variance to the $\text{EIV}_t$ over time in Figure~\ref{fig:size_leverage}. From the figure, we observe a clear trend.   The gap between the marginal variance at time $t$ and $\text{EIV}_t$ grows over time for small firms but is nearly zero for large firms. This implies that the interaction effect is stronger for small firms than larger firms. 

For a more quantitative understanding, we also calculate the summary statistics. Define $\text{RIV}_t$ as the proportion of the $\text{EMIV}_t$ to the marginal variance $\text{Var}[X_t]$,
\begin{align*}
\text{RIV}_t = \frac{\text{Var}[X_t] - \text{EIV}_t}{\text{Var}[X_t]}.
\end{align*}
The summary statistics at the 25-day time horizon are given in Table~\ref{tab:size_stats}. From the first column, it is observed that the $\text{EIV}_t$ is larger for small firms than large firms, implying that at a short horizon, small firms are more volatile than large firms, which is not surprising. From the second column, we see that the interaction effect is stronger for small firms than large firms. In particular, the annualized marginal variance roughly doubles for the smallest firms ($\text{Var}[X_t] = (1-0.51)^{-1} \text{EIV}_t$), but barely changes for largest firms at the 25-day horizon.

\begin{table}[H]
\centering
\caption{Summary statistics of the interaction effect for the portfolios constructed by size}
\begin{tabular}{ ccccc } 
\hline
Portfolios & $\text{EIV}_{25}$ & $\text{RIV}_{25}$ \\ \hline
$0\%-20\%$ & 0.037 & 0.51 \\ \hline 
$20\%-40\%$ & 0.035 & 0.41 \\ \hline
$40\%-60\%$ & 0.032 & 0.36 \\ \hline
$60\%-80\%$ & 0.030 & 0.27 \\ \hline
$80\%-100\%$ & 0.028 & 0.053 \\ \hline
\end{tabular}
\label{tab:size_stats}
\end{table}

We also check the mean-reversion effect separately for size portfolios. Recall at a short horizon, from \eqref{eq:return_asymp_kur}, the excess kurtosis by moments is $3\frac{\text{Var}[V_0]}{\mathbb{E}^2[V_0]}$. Hence, at a short horizon, the excess kurtosis measures the volatility of the variance process $V_t$. The excess kurtosis is closely related to the mean-reversion effect: if the variance process $V_t$ is not volatile or the mean-reversion effect is strong, the excess kurtosis will be small.

Unfortunately, as pointed out by \cite{kim2004more} and \cite{bonato2011robust}, the measurement of kurtosis by moments is not robust for stock returns. As an alternative, we use the quantile-based measurement developed in \cite{crow1967robust}. The centered coefficient is 
\begin{align} \label{eq:kur_cs}
\text{Excess Kurtosis}_{CS} = \frac{F^{-1}(0.975) - F^{-1}(0.025)}{F^{-1}(0.75)-F^{-1}(0.25)} - 2.91,
\end{align}
where $F$ is the empirical cumulative distribution function.  Since we have approximately 24,000 daily observations, the 0.025 and 0.975 quantiles can be measured with reasonable accuracy.  This formula is applied to size portfolios in Table \ref{tab:size_moments}. The results indicate that for small firms, in addition to the price dynamics, the variance process $V_t$ is also more volatile than for large firms.  This also implies that the mean-reversion effect is weaker for small firms than for large firms and therefore helps to explain why the interaction effect for is stronger for small firms than for large firms.

\begin{table}[H]
\centering
\caption{Summary statistics of the marginal variance, skewness, and kurtosis for the portfolios constructed by size at 1-day horizon}
\begin{tabular}{ ccccc } 
\hline
Portfolios & Annualized marginal variance & $\text{Skewness}_H$ & $\text{Excess Kurtosis}_{CS}$ \\ \hline
$0\%-20\%$ & 0.039 & -0.12 & 2.22 \\ \hline 
$20\%-40\%$ & 0.036 & -0.11 & 2.04 \\ \hline
$40\%-60\%$ & 0.034 & -0.11 & 1.96 \\ \hline
$60\%-80\%$ & 0.032 & -0.093 & 1.89 \\ \hline
$80\%-100\%$ & 0.029 & -0.053 & 1.73 \\ \hline
\end{tabular}
\label{tab:size_moments}
\end{table}

We also calculate the skewness using quantiles. We use the measurement by \cite{hinkley1975power}
\begin{align}
\text{Skewness}_H = \frac{F^{-1}(1-\alpha) + F^{-1}(\alpha) - 2F^{-1}(0.5)}{F^{-1}(1-\alpha)-F^{-1}(\alpha)}
\end{align}
with the choice of $\alpha=0.05$.  
The result is given in Table \ref{tab:size_moments}. The result indicates that the return distribution is more negatively skewed for small firms than large firms.

\section{Disentangling the puzzle for S$\&$P 500}

In this section, we explain the observed puzzle that the fitting of the S$\&$P 500 is insensitive to the correlation parameter $\rho$, even though the empirical data indicates that there is a strong leverage effect $\rho \sim -0.77$.

First, we want to confirm that the impact on the return distribution is indeed weak. To do so, we conduct an experiment. Recall that if there is no leverage effect, the marginal density follows the mixture Gaussian distribution \eqref{eq:return_cond_indep}. We show that we can fit a stochastic volatility model with no leverage effect well to the empirical data. We will rely on the generalized hyperbolic (GH) distribution (see, e.g., \citealp{eberlein1995hyperbolic,bensaida2016highly,barndorff1977infinite,barndorff1977exponentially}), which has proved to be very powerful empirically. If $Y$ follows a generalized hyperbolic distribution we write
\begin{align*}
Y \sim H(\lambda,\alpha,\beta,\delta,\mu).
\end{align*}
The PDF of a GH distribution is given by 
\begin{align*}
p_Y(x) = \frac{(\gamma/\delta)^{\lambda}}{\sqrt{2\pi} K_{\lambda}(\delta \gamma)} \frac{K_{\lambda-\frac{1}{2}}(\alpha \sqrt{\delta^2+(x-\mu)^2})}{(\sqrt{\delta^2+(x-\mu)^2}/\alpha)^{\beta-\lambda}} e^{\frac{1}{2}(x-\mu)},
\end{align*}
where $\gamma^2 = \alpha^2 - \beta^2$, and $K_{\lambda}$ is the modified Bessel function of the third kind with index $\lambda$. The parameter domain for the class of GH distributions is given by
\begin{align*}
& \delta \geq 0, \ \alpha > 0, \ \alpha^2>\beta^2, \ \text{if} \ \lambda > 0, \\
& \delta > 0, \ \alpha > 0, \ \alpha^2 > \beta^2, \ \text{if} \ \lambda = 0, \\
& \delta > 0, \ \alpha \geq 0, \ \alpha^2 \geq \beta^2, \ \text{if} \ \lambda < 0.
\end{align*}
The generalized inverse Gaussian (GIG) distribution (see \citealp{seshadri2004halphen}) is closely related to GH. If $V$ follows a GIG distribution we write
\begin{align}
V \sim GIG(\lambda,\delta,\gamma).
\end{align}
The PDF of a GIG distribution is given by 
\begin{align}
p_V(x) = \frac{(\gamma/\delta)^{\lambda}}{2K_{\lambda}(\delta \gamma)} x^{\lambda-1} e^{-\frac{1}{2} \left( \delta^2 x^{-1} + \gamma^2 x \right)}, \ x>0.
\end{align}
The parameter domain for the class of GIG distributions is given by 
\begin{align*}
\delta>0, \ \gamma \geq 0, \ \text{if} \ \lambda<0, \\
\delta>0, \ \gamma>0, \ \text{if} \ \lambda=0, \\
\delta \geq 0, \ \gamma>0, \ \text{if} \ \lambda>0.
\end{align*}
The GH distribution was originally derived in \cite{barndorff1977exponentially}; it is a Gaussian variance-mean mixture where the mixing distribution is GIG. In other words, if
\begin{align*}
Y|V = v \sim \mathcal{N}(\mu+\beta v, v),
\end{align*}
and $V \sim \text{GIG}(\lambda,\delta,\lambda)$, then the marginal distribution of $Y$ will be GH, $Y \sim H(\lambda,\alpha, \beta, \delta,\mu)$, where $\alpha^2 = \beta^2 + \gamma^2$. To connect the GIG to the SDE, \cite{sorensen1997exponential} considers the SDE
\begin{align}
dV_t = \left( \beta_1 V_t^{2\alpha-1} - \beta_2 V_t^{2\alpha} + \beta_3 V_t^{2(\alpha-1)}  \right) \ dt + \kappa V_t^{\alpha} \ dW_t,
\end{align}
where $\beta_1 = \frac{1}{2} \kappa^2(\lambda-1) + \kappa^2 \alpha, \ \ \ \beta_2 = \frac{1}{4} (\kappa \gamma)^2, \ \ \ \beta_3 = \frac{1}{4} (\kappa \delta)^2$. This SDE has a stationary distribution which is GIG. Note that if $\alpha = \frac{1}{2}$, then the diffusion process is the solution to 
\begin{align}
dV_t = \left( \beta_1 - \beta_2 V_t +  \frac{\beta_3}{V_t} \right) \ dt + \kappa \sqrt{V_t} \ dW_t,
\end{align}
which is the CIR-process with an additional $\frac{\beta_3}{V_t}$ in the drift term. Empirically, the volatility of the market is never zero. In the CIR-process, the variance can become zero if the Feller condition is violated, and imposing the Feller condition limits our ability to calibrate the model to data.  The additional $\frac{\beta_3}{V_t}$ term prevents the variance from becoming zero, while leaving more freedom to fit the data.

\begin{figure}[H]
\centering
\includegraphics[scale=0.6]{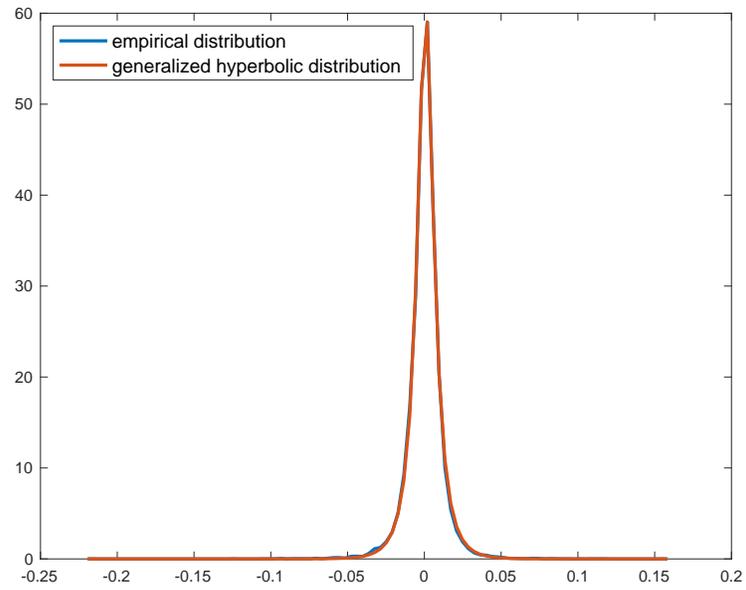}
\caption{The fitness of the S$\&$P 500 by the generalized hyperbolic distribution at the daily scale}
\label{fig:SP500_fit}
\end{figure}

The fitting of the empirical distribution of the S$\&$P 500 by GH is shown in Figure~\ref{fig:SP500_fit}. Visually, the fit is virtually perfect. Quantitatively, we apply the Kolmorogov-Smirnov test (see, e.g., \citealp{weiss1978modification,chicheportiche2011goodness}) to verify with daily returns drawn one-month apart so that the data is only very weakly dependent. The test fails to reject the null hypothesis that the sample is drawn from the GH distribution at the $5\%$ significance level, confirming that we have a good fit. This experiment, which shows that it is possible to fit the S$\&$P 500 return distribution with the GH distribution, which is derived under the assumption of zero leverage effect, confirms that the leverage effect has only a very weak impact on the S$\&$P 500 return distribution.

\begin{figure}[H]
\centering
\includegraphics[scale=0.6]{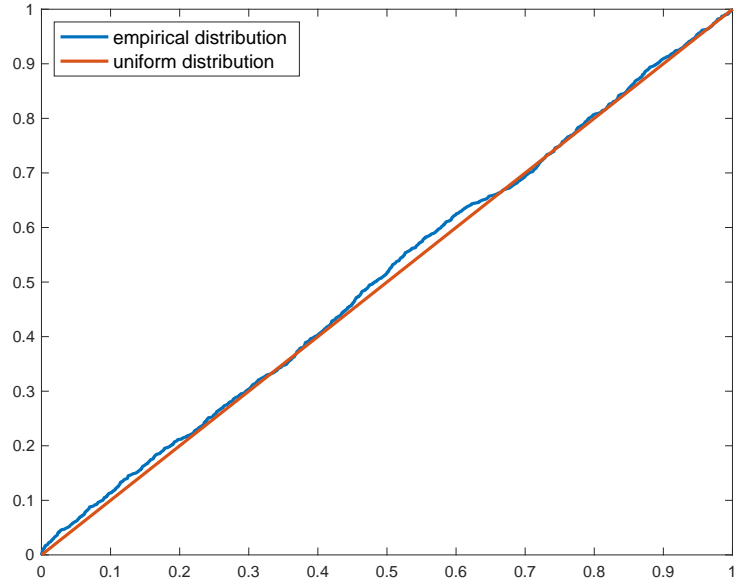}
\caption{The Kolmorogov-Smirnov test test of the S$\&$P 500 at the daily scale}
\label{fig:SP500_kstest}
\end{figure}

\begin{figure}[H]
\centering
\includegraphics[scale=0.6]{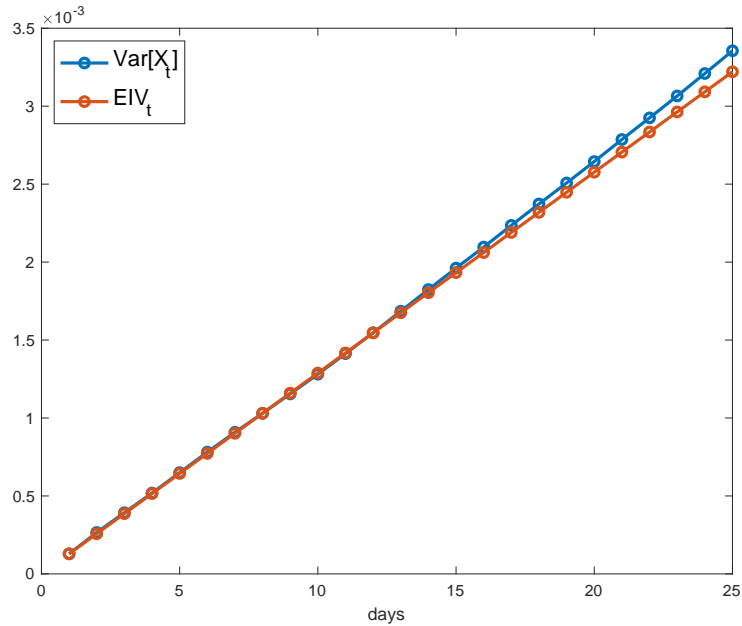}
\caption{The interaction effect of the S$\&$P 500 over time}
\label{fig:SP500_leverage}
\end{figure}

We now turn to the interaction effect for the S$\&$P 500. The comparison of the dynamics of marginal variances with $\text{EIV}_t$ is plotted in Figure~\ref{fig:SP500_leverage}. The marginal variance is quite close to $\text{EIV}_t$, indicating that the interaction effect is quite weak. We believe this weak interaction effect is the reason that the leverage effect has little impact on the distribution for S$\&$P 500.

To back up our story, we check the mean-reversion property of the S$\&$P 500. We found that the interaction effect is weak for the S$\&$P 500. From the literature, the leverage effect is strong. By our analysis, this implies that the mean-reversion effect for S$\&$P 500 must be strong. The excess kurtosis, calculated using Equation \eqref{eq:kur_cs} of S$\&$P 500 is relatively low, at 1.72, which confirms that the mean-reversion effect is strong, consistent with our analysis.\footnote{The result is quite similar to that for large firms, 1.73.  This is not surprising since the S$\&$P 500 is composed essentially of the 500 largest firms by market capitalization.}

\section{Conclusion}

In this paper, we study the impact of the leverage effect on the return distribution. In particular, we focus on the marginal distribution since it can be directly formed from the empirical data. We find that the leverage effect is important, but it need not have a big impact on the return distribution. A strong leverage effect can be negated by a strong mean-reversion effect. Hence, when studying the return distribution properties, one must consider the interaction effect, which is the mixture of the leverage effect and the mean-reversion effect. 

The interaction effect is not directly measurable. Our measurement relies on the dynamics of the marginal variance, which can be estimated accurately. A direct impact of our measurement is that the annualized marginal variance of the return will grow over time until it converges for a strong interaction effect, but will barely change for a weak interaction effect. The study of the interaction effect also shed light on the performance of the SRTR. Even with a weak mean-reversion effect, if the interaction effect is strong, the SRTR will not give an accurate approximation.

When applying our methodology to empirical data, we observed some interesting phenomena.  We found that the interaction effect is stronger for small firms than large firms. We resolved the puzzle that fitting the S$\&$P 500 return distribution is insensitive to the leverage effect. By employing the GH distribution, a marginal distribution for a class of stochastic volatility model with no leverage effect, we confirmed that the S$\&$P 500 return distribution can be fitted well without the leverage effect. By our method, we found the interaction effect for S$\&$P 500 is weak, which we believe is the explanation to the puzzle.

\section*{Acknowledgments}
We thank Robert Anderson and Lisa Goldberg for helpful discussions and comments.


\bibliography{Manuscript}

\end{document}